# Machine learning prediction of COVID-19 severity levels from salivaomics data

Aaron Wang[1, Y ‡], Feng Li[1], Samantha Chiang[1], Jennifer Fulcher[2], Otto Yang[2], David Wong[1], Fang Wei[1, Y]

**1** School of Dentistry, University of California, Los Angeles, CA, USA
**2** Department of Infectious Diseases, David Geffen School of Medicine, University of California, Los Angeles, CA, USA

Y These authors contributed equally to this work.
‡ This author contributed to this work as a student at Brown University.

## Abstract

The clinical spectrum of severe acute respiratory syndrome coronavirus 2 (SARS-CoV-2), the strain of coronavirus that caused the COVID-19 pandemic, is broad, extending from asymptomatic infection to severe immunopulmonary reactions that, if not categorized properly, may be life-threatening. Researchers rate COVID-19 patients on a scale from 1 to 8 according to the severity level of COVID-19, 1 being healthy and 8 being extremely sick, based on a multitude of factors including number of clinic visits, days since the first sign of symptoms, and more. However, there are two issues with the current state of severity level designation. Firstly, there exists variation among researchers in determining these patient scores, which may lead to improper treatment. Secondly, researchers use a variety of metrics to determine patient severity level, including metrics involving plasma collection that require invasive procedures. This project aims to remedy both issues by introducing a machine learning framework that unifies severity level designations based on noninvasive saliva biomarkers. Our results show that we can successfully use machine learning on salivaomics data to predict the severity level of COVID-19 patients, indicating the presence of viral load using saliva biomarkers.

## Introduction

Saliva is a non-invasively available biofluid that contains many types of readily detectable biomolecules, including antibodies, and other classes of biomarkers that have clinical diagnostic properties.
   Furthermore, existing scientific studies have shown that salivary diagnostics for COVID-19 are effective for screening and managing COVID-19. For patients, self-collection of non-invasive biofluids has proved to be more convenient and user-friendly. As researchers, we would like to be able to use the salivaomics data provided by the biomarkers in patient saliva to make meaningful inferences. A machine learning model that analyzes biomarker data in saliva to predict patient severity would address both.

### COVID-19

The COVID-19 pandemic, since March 2020, has brought to the forefront the urgency and necessity of testing for SARS-CoV-2 infection (mRNA), viral load as a measure of infectivity (antigen), and immunity (antibodies). At this time there are 200+ EUA-approved molecular tests for SARS-CoV-2 RNA, 14 EUA-approved SARS-CoV-2 antigen/infectivity tests, and 67 EUA-approved serology tests. All these tests are single-plexed, conveying a single dimension of COVID-19 in an individual. While the pressing need is to know if there is infection by the pathogen (gRNA) and the infectivity (antigen), the start of population vaccination quickly elevated the need to assess host immunity development (antibodies). The ability to assess infection/re-infection, infectivity, and immunity to SARS-CoV-2 in any individual is of increasing clinical relevance as breakthrough infections and interest in resuming normal life make knowing the COVID-19 status of any individual a matter of pressing importance.

## EFIRM

Electric field–induced release and measurement (EFIRM) is an electrochemical biosensor platform that has been extensively peer-reviewed and has demonstrated clinical utility across a variety of disease systems, including oral cancer, early-stage lung cancer, and autoimmune disorders [1–3]. Salient features of the platform include its ability to sensitively and robustly detect a wide class of omics targets/biomarkers (RNA, proteomic, and genomic) in saliva. EFIRM is a direct-detection plate-based assay that utilizes electrochemistry to capture target molecules directly from a minimal volume of biofluid (50 μL), and has been shown to be a robust platform for the direct detection of biomarkers in basic science and translational research on genomic, proteomic, and small molecular targets [4–6]. EFIRM is optimized for the direct detection of salivary omics targets without sample processing, with a performance that exceeds current EUA assays using PCR and ELISA/Luminex as reference technologies. The COVID-19 pandemic catalyzed the need for saliva-based detection of SARS-CoV-2 and has driven our recent advancement of EFIRM technology for the concurrent detection in saliva of SARS-CoV-2 mRNA (infection), antigen (infectivity), and antibodies (immunity) with a high level of performance. This performance level can immediately benefit researchers and patients in the midst of the COVID-19 pandemic.

## Machine Learning

Machine learning (ML) is a branch of artificial intelligence (AI) where computer systems are able to automatically learn and improve from patterns and observations in data without being explicitly programmed [12]. This allows computers to mimic a human's ability to learn without assistance and adjust their actions accordingly. In general, there are three types of learning that ML algorithms are capable of: supervised, unsupervised, and reinforced. Supervised learning requires that the data the model is trained on be labeled, such that the model will learn to approximate the relationship between samples and their labels. Unsupervised learning is performed on unlabeled data, in which the model can independently identify patterns, clusters, or other organization without previous classification. Reinforcement learning is a behavior-based learning: an autonomous agent learns about its interactive environment through a programmed rewards system. We use supervised learning in this work because our data is well-labeled, with each patient classified in regard to an existing severity level.

Within ML, deep learning (DL) is a specific family of models that rely on artificial neural networks, an algorithmic system of artificial neurons that processes information and learns complex patterns in data and is broadly inspired by biological systems in the human brain. DL has largely been responsible for the recent advancements in artificial intelligence in computer vision, natural language processing, and reinforcement learning.

In medicine, DL continues to be crucial in a variety of research and clinical applications, including medical imaging diagnosis and genomic data analysis. For example, brain tumor segmentation on patient MRI data has been useful in automating the glioblastoma segmentation process and in minimizing variance in comparison with radiologists' segmentations performed by hand [13]. In the field of genomics, DL has been used to encode histone modification data in order to predict gene expression [14]. Our work adds to the applications of DL in medicine by exploring the viability of ML and DL to learn from salivaomics data in predicting COVID-19 patient severity levels.

## Material and Methods

### Data Processing

The clinical cohort consisted of acutely infected COVID-19 patients, recovered COVID-19 patients, and healthy control patients without vaccination, with sample sizes of 41, 74, and 50, respectively, for a total of 165 individuals. Each patient had the following biomarkers measured, along with his or her number of visits to his or her physician, number of days since first developing symptoms, and severity level: saliva anti-RBD (SARS-CoV-2 receptor-binding domain) IgG (ng/mL), serum anti-RBD IgG (ng/mL), saliva anti-RBD IgA (ng/mL), serum anti-RBD IgA (ng/mL), saliva anti-RBD IgM (ng/mL), serum anti-RBD IgM (ng/mL), EFIRM saliva SARS-CoV-2 (N) antigen TCID50/mL, and EFIRM saliva SARS-CoV-2 gRNA (N2, NL) LAMP RNA (qualitative). Samples were collected by whole saliva and nasal swab for molecular testing (FDA-approved DiaSorin Simplexa platform). Each patient was also given a severity level designation from NIH clinical trials. Although the designations formally range from 1 to 8, the provided data include values of only 1 to 6, inclusive. We used Pandas DataFrames to load and process the raw data, remove any incomplete samples, and split the features from the severity-level labels of the patients. Healthy and recovered patients were not given a severity level—thus, in order to use these patients' measurements for model training and evaluation, we set their labels to 1, the lowest possible designation, so that they could still be used for supervised training. Including all individuals in our experiments gave our models a larger amount of data to learn from, a characteristic from which every machine learning model benefits greatly. We show the class imbalance this created in Figure 1.

With regard to feature selection, all features were kept for the developed models for the sake of continuity. Furthermore, by including all measured salivaomics features in the input, we avoided any possible omitted variable bias. We also took a more rigorous approach to support our inclusion of all features by building an extra-trees classifier for feature selection. Extremely Randomized Trees Classifier (Extra-Trees Classifier) is an ensemble learning technique that aggregates the results of multiple collected de-correlated decision trees to output its classification result. During the construction of the forest, for each feature the normalized total reduction used in the feature split decision is computed. The result is known as the Gini importance of the feature and can be used to describe how important the feature is to the dependent variable. After running the Extra-Trees Classifier on our features in Python using the Sci-Kit Learn package, we found that the Gini importance values of the features were 0.06, 0.10, 0.06, 0.07, 0.08, 0.07, 0.07, 0.08, 0.26, and 0.15. It should be noted that all non-EFIRM values, whether serum or saliva-based, clustered around the .07 importance level, indicating that no particular measurement was superior to another in indicating severity level. With regard to the EFIRM measurements, the serum and saliva-based EFIRM measurements posted similarly high levels of Gini importance, and thus both were naturally included in the experiments as well.

**Fig 1. Frequency Distribution of Severity Levels** The distribution of severity levels is right-skewed after data augmentation to include the healthy individuals, leading to a class imbalance.

### Baseline Model

Ordinary Least Squares (OLS) regression is used to model the relationship between explanatory variables and a response variable by fitting a linear equation to observed data, minimizing the sum of squared differences between the predicted and ground truth data. We began looking at the use of OLS in order to establish the possibility of using advanced ML techniques on salivaomics data as well as to compare ML with the neural network model. Due to the nature of OLS, the baseline model yields floating-point numbers as predictions, despite the fact that severity levels are strictly integers; hence, predictions are rounded in order to align them with the clinical task. We implemented our baseline OLS model with Sci-Kit Learn in Python.

### Neural Network Architecture

Feed-forward neural networks, also known as multilayer perceptrons, are a type of advanced ML computing system able to analyze and learn patterns in data. First proposed in 1944 by Warren McCullough and Walter Pitts and popularized in the 1980s, this form of ML, now known as deep learning (DL), has since been extensively used for a wide variety of applications, including self-driving cars [7], natural language processing (Google Translate) [8], and reinforcement learning (AlphaGo) [9]. In this paper, we implement a fully connected feed-forward neural network for severity-level prediction on salivaomics data using the Keras framework. Our supervised model is composed of five total stages and over 2,000 trainable parameters. There are two ways to frame the task of predicting severity level designation: single-value prediction or multiclass classification. The first task outputs a single value in the set of real numbers which represents the model's severity level prediction. This flexibility allows the model to predict floating points to minimize the difference between predicted and observed values, but is not realistic as observed values are only integers. The second task assigns probabilities to each possible output, from 1 to 6, representing the model's confidence in each output being the correct value, and the value with the highest probability is chosen as the model's final output. This way of framing the task is more realistic, as we mimic the researchers who choose a severity level given the output space of integers from 1 to 6, inclusive. Hence, we framed the task of predicting severity level as a multi-class classification problem.

### Dropout

Overfitting in ML occurs when a model fits/learns the noise in the training data to the extent that it decreases the performance of the model on unseen data. This occurs because the noise or random fluctuations in the training data are learned as important patterns by the model. These memorized fluctuations from the training data do not apply to new data and thus negatively impact the model's ability to generalize. Generalization of a model to unseen samples is what allows ML algorithms to make powerful predictions and classify data.

Dropout is a regularization technique first used in 2014 that prevents neural networks from overfitting [10]. Dropout regularization randomly drops neurons from the neural network during training in each iteration. During training, dropout samples form several distinct "thinned" networks, similar in notion to the actions of ensemble methods like bagging and boosting in classical ML [11]. At test time, a unified "unthinned" network approximates the effects of averaging the predictions of the thinned networks, significantly reducing overfitting [10]. In dropout, the network cannot rely on one feature but rather has to learn which robust features are useful. We used dropout with a rate of 0.1 after the model's first layers as a mode of regularization to complement the large number of neurons in the initial layer.

## Results

### OLS Regression

Our baseline model performed as expected, achieving a mid-range accuracy of 0.55 on the validation set. This shows that saliva contains SARS-CoV-2 omics content that can be predictive of the severity of a viral disease such as COVID-19. In Figure 2, we plot the predicted severities against the ground-truth labels from the validation set.

**Fig 2. Scatter Plot of OLS Regression, True and Predicted Severity** The points are close to the regressed diagonal line with a wider confidence band. For lower severity levels, the OLS regression model tends to overestimate, and for higher severity levels the model tends to underestimate, indicating that this particular model has high bias and is unable to fit the data properly.

Our hypothesis that ML can be used on salivaomics data is supported by the linear regression's normally distributed residual plot in Figure 3A. A mean absolute error of 0.70 was achieved, showing that the model was, on average, off by 0.70 when scoring severity. This model also achieved a root mean square error of 1.04.

**Fig 3. Residual Plots of OLS Regression and Neural Network** Both residual distributions are normally distributed, indicating the efficacy of learning for these two models. Note, however, the greater center density for the neural network, indicating less error than in the OLS regression.

### Neural Network

The neural network model outperformed the baseline model in all performance-based metrics. The model achieved an accuracy of .85 on the test set, beating the baseline model by .30. We can conclude from this that not only can saliva be predictive of the severity of a viral disease such as COVID-19, but that the patterns extracted from the salivaomics data are complex enough to require a neural network model to learn them. Indeed, this means that a more complex model outside of our neural network may prove to be even more effective at learning from salivaomics data. This is further supported by an analysis of the neural network's residual plot, which shows that the distribution of residuals is much more centered at zero with a smaller variance than is the residual distribution of the baseline model. Table 1 displays the overall precision-recall metrics for each severity level. For a severity level of 1, we see an impressive F1 accuracy score of .96, which means that the model is well able to classify healthy patients. For severity levels 2–4, we see F1 scores of zero, indicating that the model was unable to classify
**Table 1. Precision Recall By Severity Level**

|   | Precision | Recall | F1 Score | Samples |
|---|-----------|--------|----------|---------|
| 1 | 0.93 | 1.00 | 0.96 | 27 |
| 2 | 0.00 | 0.00 | 0.00 | 1 |
| 3 | 0.00 | 0.00 | 0.00 | 1 |
| 4 | 0.00 | 0.00 | 0.00 | 1 |
| 5 | 0.50 | 0.33 | 0.40 | 3 |
| 6 | 0.00 | 0.00 | 0.00 | 0 |

Outlier values are reviewed in Discussion.

these results well. The reasoning behind this is that for each of these severity levels there was only one patient in the test split with that severity level; hence, if the model incorrectly labels it once, the F1 score for this particular severity level is zero. For severity level 5, the test split has 3 individuals and the model is able to correctly identify the severity level of one of these patients; thus, we see a precision of 0.50 and a recall of 0.33. For severity level 6, because there were so few patients with this severity level, all patients with this severity level were randomly split into the training set. Because the model did not evaluate on this class, its F1 score is zero.

## Discussion

We have successfully shown the deployment of a novel ML framework on salivaomics data to predict COVID-19 severity level. We should note, however, some key insights into our data that limited the function of our model. The number of samples available at the time of this publication greatly limited the model's ability to learn about the relationship between the biomarker levels and severity designations. With only 1 patient with a severity level of 2, 2 patients with a severity level of 3, 9 patients with a severity level of 4, 14 patients with a severity level of 5, and 3 patients with a severity level of 6, finding a train/test split to balance the number of data points for the model to train on while having enough data points to evaluate the model's performance was difficult. For example, because there was only one patient with a severity level of 2, it was difficult to decide if the model should train on this patient or if we should evaluate the model on its ability to extrapolate from other severity level data. The final decision was to remain hands-off and randomize the train/test split with a well-standardized random seed of 42. Future work could build off of this hands-off approach by carefully and actively separating samples into training and test sets based on the distribution of the data available.

 In an attempt to alleviate the issue of a lack of data, we chose to use the data of healthy patients by setting their severity levels to one and using them as samples for our model. By giving more data to our model to work with, we believed that the model would better understand what biomarker levels characterized a healthy patient and thus would be able to learn the patterns behind what biomarker levels characterized different tiers of sick patients. Our approach accomplished this. When the healthy-patient data points were not included in the experiments, the mean absolute error of the test set on the neural network model was 1.33, as compared with 0.36 when the healthy patients were included. Furthermore, the model accuracy when these data points were not included was a mere 0.44, as opposed to 0.85 when the data points were included.

However, this approach is not without drawbacks. This data augmentation approach induces class imbalance that biases the results of the model. Despite the headline model accuracy of 0.85, this metric is skewed by the overwhelming well-classified class of severity level 1. If we were to weight the F1 accuracies of each class equally, as opposed to weighting them based on the frequency of each class, we would find that the macro-average accuracy of the model is 0.23 with the healthy individuals included. This is in comparison with weighting the class accuracies based on frequency, from which the accuracy is 0.83. Had we not included the additional healthy patients, the macro-average accuracy of the model would be 0.31, 7 percentage points higher than if we had included the healthy cohort, and the weighted average would be 0.39. These results show that although the addition of the healthy cohort's information helped our model to perform well, it was not without penalty. Our model sacrifices the correct classification of the minority (the COVID-19 patients) by guessing more towards the majority (healthy) class when unsure. In other words, if the model is unsure of the output, it is more likely to assume the individual is healthy because the majority of people it has seen before have been healthy. Future work can build on this conclusion by oversampling patients with higher severity levels or, if that is not possible, creating a data synthesis pipeline that mimics the sampling of these patients.

## Conclusion

In this study, we propose a machine learning framework that can accurately predict COVID-19 severity levels from salivaomics data. Our findings not only validate previous observations of the presence of viral load in saliva, but also provide new insights into the prospects of the novel use of machine learning on salivaomics data. Our neural network model not only achieves state-of-the-art accuracy of 0.85 on COVID-19 patients, but also is exceptionally computationally efficient and can be useful in a clinical setting, as a single inference takes only 40 milliseconds of CPU time. Further studies could be performed using more complex deep learning models on salivaomics data in predicting diseases other than COVID-19.

## Acknowledgments

Funded by R25 DE030117, U18 TR003778, U54 HL119893.

## Conflicts of Interest

David Wong has equity in Liquid Diagnostics LLC and RNAmeTRIX.

Figure 1	Click here to access/download;Figure;Figure 1.tiff

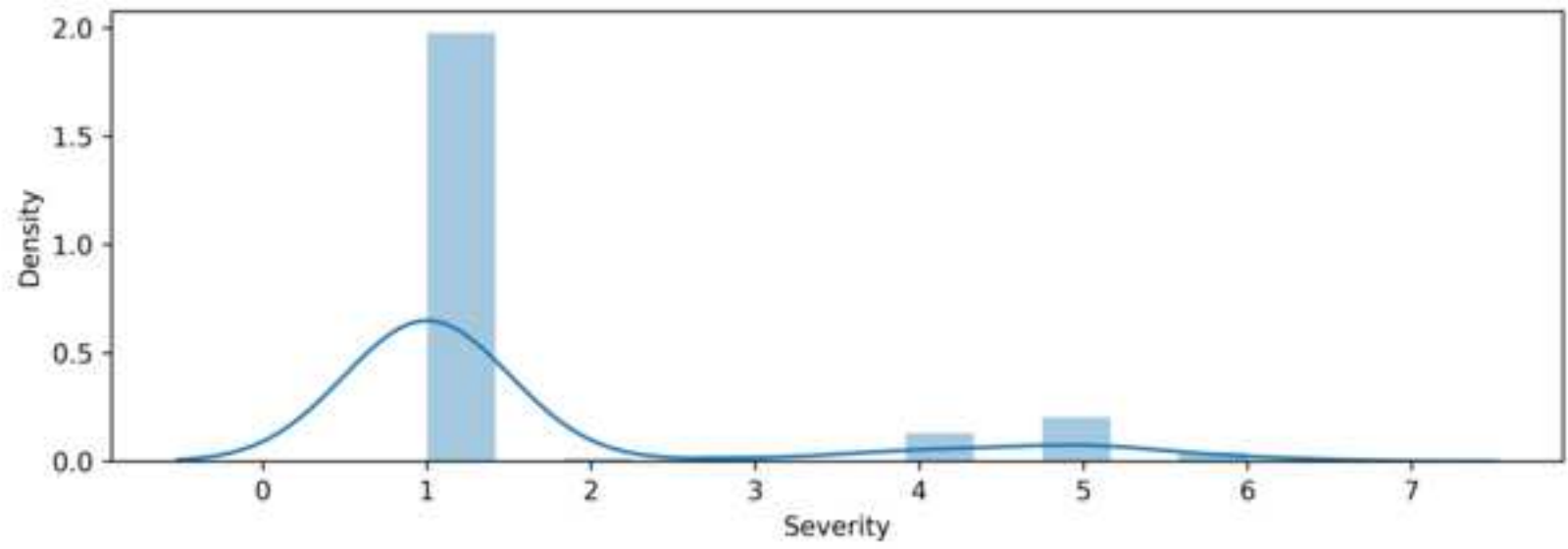

Figure 2							Click here to access/download;Figure;Figure 2.tiff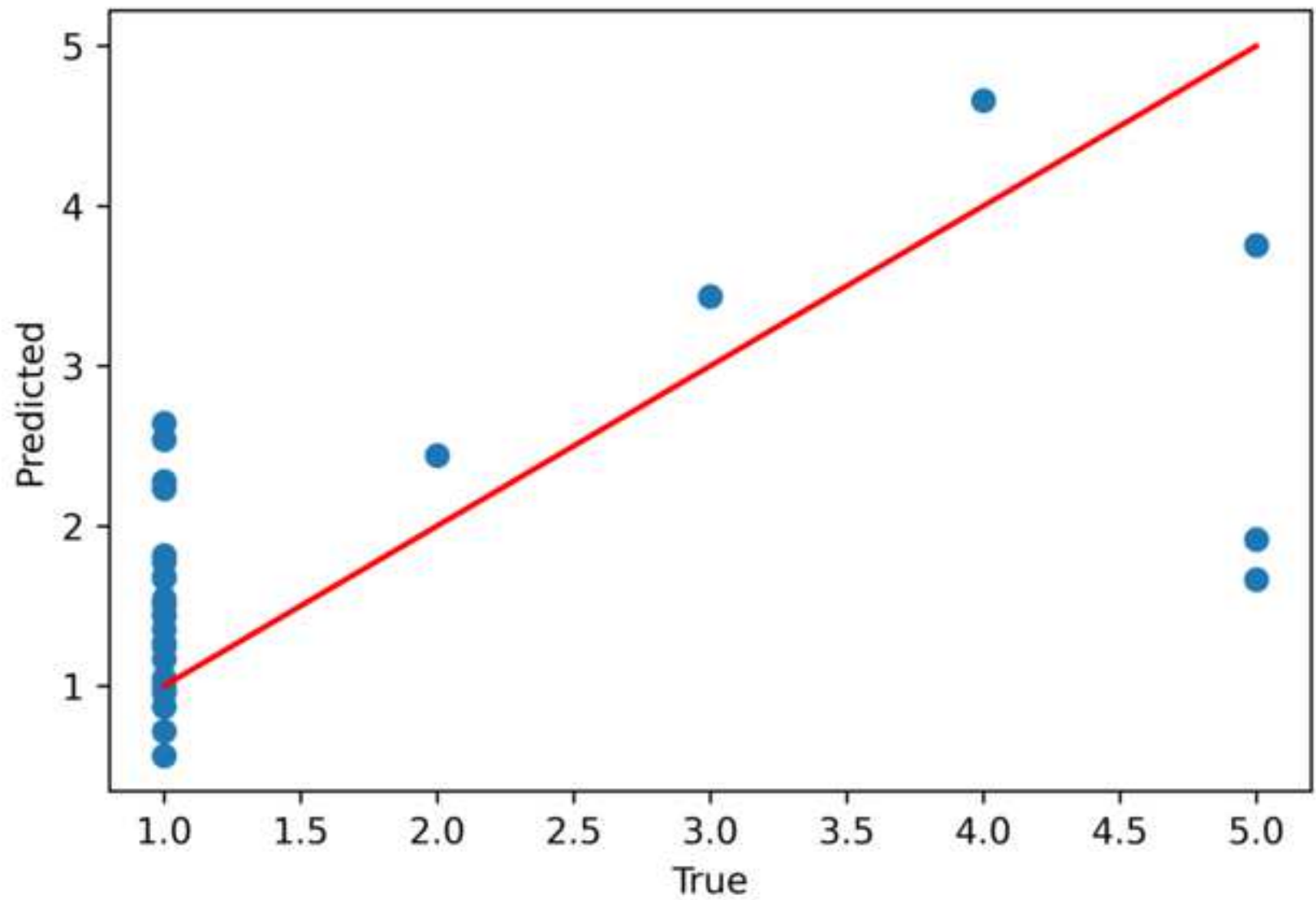

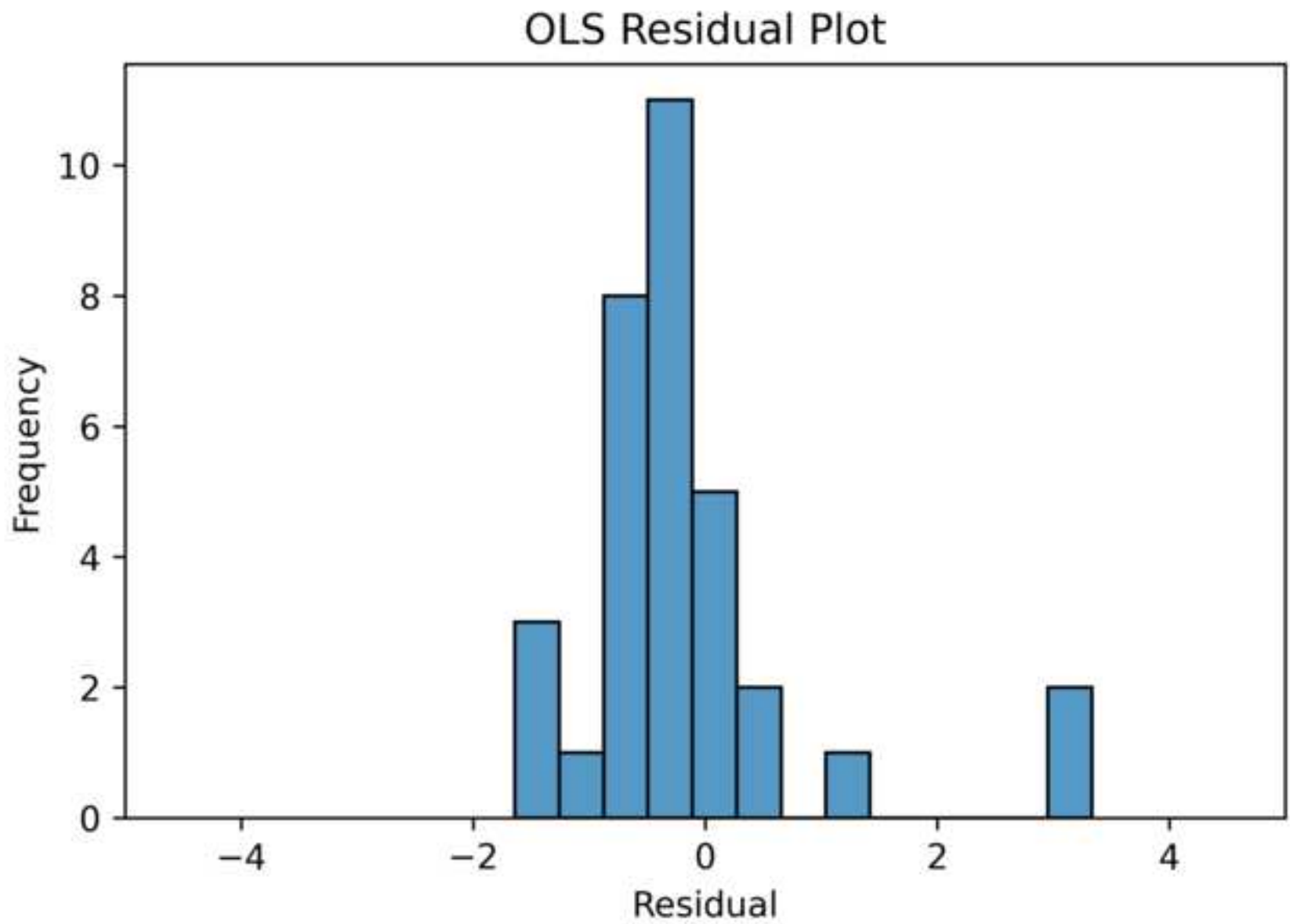

Figure 4    Click here to access/download;Figure;Figure 3b.tiff

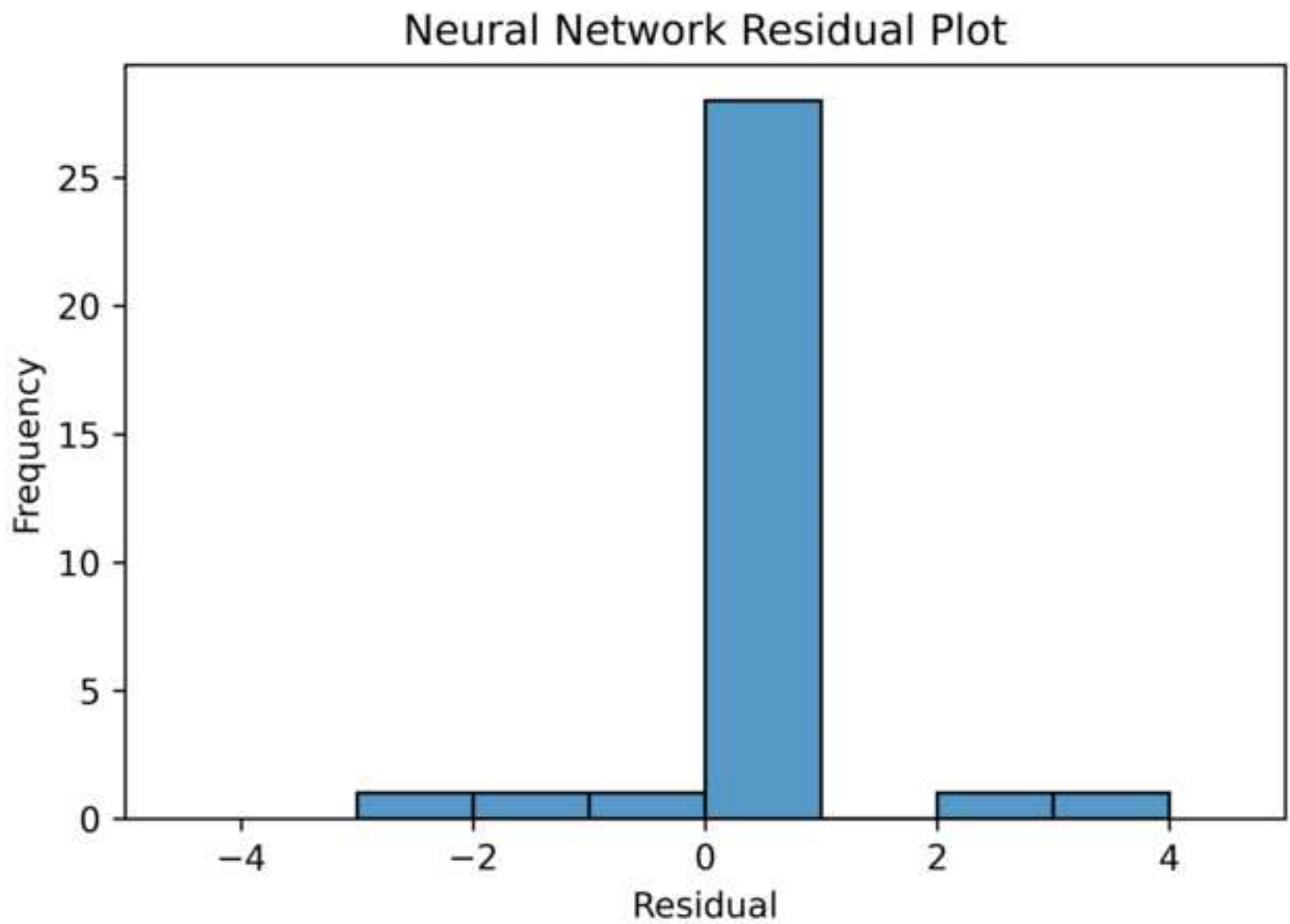